\documentclass[prb,twocolumn]{revtex4}
\usepackage{graphicx}
\usepackage{epsfig}
\usepackage{amssymb,latexsym,amsmath}

\usepackage{hyperref}
\usepackage{soul}

\usepackage{ulem}
\usepackage{xcolor}

\begin{document}

\title{ \Large{ \bf A new angle on the well-known FMO photosynthetic complex }}

\author{A.-M. Dar\'e}\email{Anne-Marie.Dare@univ-amu.fr}
\author{C. Demarez}%\email{Cyriel.Demarez@etu.univ-amu.fr}
\author{J. Missirian}%\email{Justine.Missirian@etu.univ-amu.fr}
\author{F. Michelini}%\email{Fabienne.Michelini@univ-amu.fr}
\affiliation{Aix Marseille Univ, Universit\'e de Toulon, CNRS, IM2NP, Marseille, France \\
Facult\'e des Sciences de St J\'er\^ome, Case 142, 13397 Marseille Cedex 20, France}
\date{\today}
\begin{abstract}
Using a formalism adapted to study transport in quantum open systems, that is the non-equilibrium Green's function formalism, we revisit the working principle of the most popular photosynthetic complex, namely the Fenna Matthews-Olson complex. We underline the driving force behind the exciton flow. 
We also show that in a realistic parameter regime, two mechanisms, one of which has not yet been proposed, may be at work to protect the reaction center from overheating.  
\end{abstract}

\maketitle

\section{Introduction}

The Fenna-Matthews-Olson complex (FMO) [\onlinecite{Olson62,Fenna75}]  is a sunlight harvesting system that can be found in green sulfur bacteria living in low illumination conditions.  
It is among the smallest and simplest pigment-protein complexes for light-harvesting appearing in nature, and as such, has been the subject of many experimental studies, as well as theoretical ones. 
The FMO transfers excitations from a light-harvesting antenna (called {\it chlorosome}) to a photochemical {\it reaction center}, where chemical reactions  take
 place, thus converting sunlight into chemical energy.
This natural nanodevice involves a trimer structure of seven bacteriochlorophylls (BChl) each, to which an eighth recently discovered peripheral BChl has been added. 
These three additional BChls in total are assumed to play a bridging role between chlorosome and FMO per se [\onlinecite{Schmidt10}], or perhaps a more subtle and counterintuitive inter-monomer task [\onlinecite{Lopez22}]. 
A protein scaffold holds the complex monomer setting. 

The possibility of quantum behavior when operating under natural conditions has aroused the interest of chemistry and physics communities, as well as the biology one, following works of Engel's group that have revealed long-time coherence succeeding short laser pulse excitations  [\onlinecite{Engel07,panitchayangkoon10}].
These long-lived coherences have been challenged and questioned  [\onlinecite{Duan17,Harush21}], and they probably 
have more a vibrational than an excitonic origin  [\onlinecite{Wilkins15}].
All these studies, with their controversies, have nevertheless significantly enhanced our understanding of widespread light harvesting devices in general, and have brought out 
a subtle picture of complexity: an interplay of classical and quantum properties [\onlinecite{Plenio08,delRey13,kramer14,Mattioni21}] is probably at the helm. 
There are still open questions about the pigment-protein complex, for example concerning the role of the eighth BChl, or concerning the sites connected to the reservoirs, 
see for instance Refs. [\onlinecite{Milder10,Chaillet20}]. 

In the present paper we adopt a slightly off-center viewpoint, seeing the photosynthetic complex as a device for quantum transport.
Making a parallel between biological and condensed-matter devices, is not new, see for instance Refs. [\onlinecite{blankenshpi11,scholes11,Mazziotti12,Alharbi15}], and 
our aim is not so much to gain a better understanding of how these efficient natural devices work, as to draw inspiration for the design of artificial devices. 

In these light-harvesting systems, energy carriers are the so-called Frenkel or local excitons. 
These carriers have recently attracted a great deal of interest in condensed mater systems, in particular in two-dimensional heterostructures [\onlinecite{Mueller18}]. 
The properties of the FMO complex are usually addressed in the Lindblad form of the Master equation [\onlinecite{Plenio08,Caruso09,Harush21}], that can study dynamics as well as steady-state properties. This formalism is constrained by the assumption of weak-coupling between the complex and its surroundings.
Beyond this perturbative hypothesis, the hierarchical equation of motion approach (HEOM) is a powerful yet numerically demanding method [\onlinecite{Ishizaki09,Chen15,Lambert23}].
The formalism we adopt, namely non-equilibrium Green's function (NEGF) formalism, has several special features and attributes, and does not suffer of the weak-coupling limitation. 
It can handle bosons as well as fermions, performing calculations in the entire respective Fock spaces. 
It is designed to study quantum transport properties in open quantum systems, and can address also the 
stationary regime as well as the time-dependent one.  We restrict the present study to the former case. It treats exactly - from the FMO side - the coupling between the FMO with its seven BChls and the input and output 
reservoirs, that is, the chlorosome and the reaction center, as well as the coupling between the BChls and the photon bath responsible for radiative decay. NEGF can in principle deal with frequency dependent exciton-vibration interaction, although this is more computationally demanding. 
In the present paper we make high temperature and vibrational low-frequency assumptions - that lightens this aspect of the calculations - to implement the exciton-vibration interaction. 

The use of NEGF in the present context is, to the best of our knowledge, limited to one study conducted by Pelzer {\it et al.} and published ten years ago [\onlinecite{Pelzer14}].
This not so well-known paper has shown that interaction with vibrations, even in an elastic manner can greatly enhance the energy transfer through the FMO by excitons, in line with environment assisted quantum 
transport mechanism (EnAQT) [\onlinecite{Plenio08,Rebentrost09,Caruso09}].
It also provided visual maps of the different transport pathways, which depend on the exciton-vibration coupling strength, likely to cause backflows. 

In the present paper we extend this pioneering work by investigating the influence of the reservoir fillings 
which entail the possibility of exciton return from the reaction center. 
We highlight an affine relation between incoming and outgoing currents and exciton populations inside the respective reservoirs. 
To deepen this finding, and understand the properties of the coefficients of this affine relation, we take the liberty to redesign the FMO architecture, revealing the significant influence of the FMO spectrum.
Finally we sketch two mechanisms, the first one not yet mentioned, that may limit the exciton flux, thus protecting the reaction center:\\
- by reversal of exciton output flow above a threshold exciton number on the reaction center followed by dissipation of the excess energy,\\
- by limitation, or even reversal, of exciton output flow through conformal reorganization.

The paper is organized as follows: 
after a detailed presentation of the model in section II, we expose the results of our calculations in section III, before concluding in section IV and supplementing the paper with appendices that provide  
detailed relations of NEGF formalism for bosons, as well as discussion about the exciton-vibration interaction.

\section{Model}

\subsection{Hamiltonian}

Along Pelzer's lines [\onlinecite{Pelzer14}], we adopt the following Hamiltonian to describe the FMO complex and its surroundings:
\begin{equation}
H = H_{FMO} +H_{rec}+H_{pt} +H_{pn}+H_{e-pn +}H_T+H_c + H_r \ ,
\label{HamiltonienTotal}
\end{equation}
with $H_{FMO}$ the part that  characterizes the FMO complex with seven chromophores labelled $i$, isolated from all the reservoirs:  
\begin{equation}
H_{FMO} = \sum_{j=1}^7 E_i d^\dagger_i d_i + \sum_{i \neq j} V_{ij} d^\dagger_i d_j \ .
\label{HFMO}
\end{equation}
\begin{figure}
\includegraphics[width=0.4\textwidth]{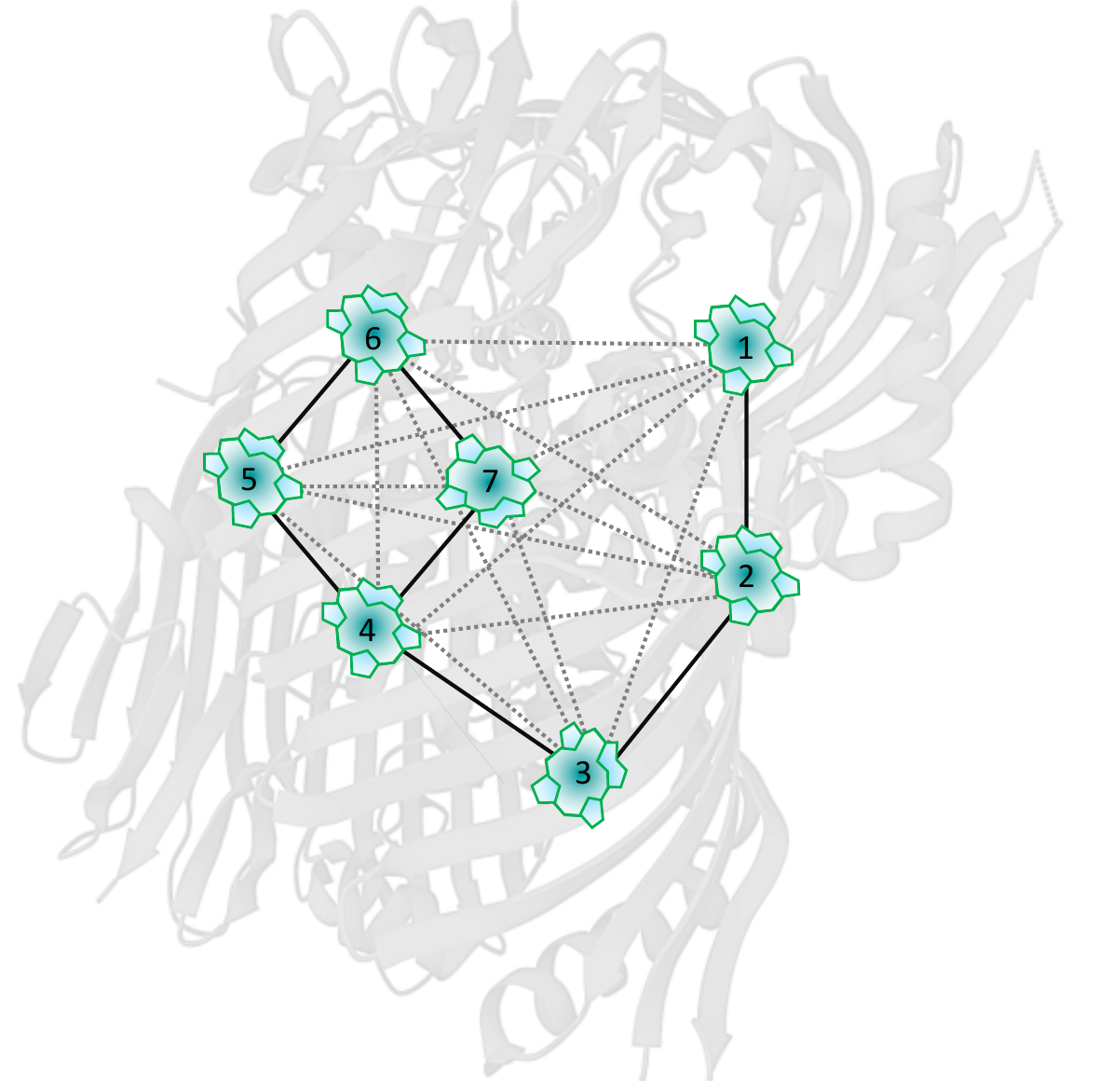}
\caption{Illustration of one monomer of the FMO complex, with standard site numbering  [\onlinecite{Fenna75}] and the surrounding protein evoked in grey. The main hopping parameters $|V_{ij}|$ are indicated in solid lines, the weakest in dotted ones. }
\label{FMObase}
\end{figure}
These seven BChl macromolecules, drawn in blue in Fig.~\ref{FMObase}, are described by a single degree of freedom: in their excited state they host an exciton of energy $E_i$. 
These energies change from site to site because of the presence of the surrounding protein. 
In the previous Hamiltonian,
$d^\dagger_i (d_i)$ creates (annihilates) an excitation at chromophore $i$ that can propagate through the $V_{ij}$ coupling terms.
$H_{rec}$ is the Hamiltonian part taking into account the radiative recombination inside the FMO complex: the exciton may annihilate thus creating a photon ($e_i^\dagger$). By hermiticity ($hc$) the reverse process can also occur:
\begin{equation}
H_{rec} = V_b \sum_{i=1}^7 (  e^\dagger_i  d_i+hc ) \ .
\end{equation}
Disadvantaging one or other of the processes is ensured by the control of the photon population onto the FMO (see later). 
$V_b$ is the optical coupling between the BChls and the local photon bath. It is assumed to be the same for all chlorophylls. 
The Hamiltonian of the local photon baths, whose energies are $E_i$, simply reads
\begin{equation}
H_{pt} = \sum_{i=1}^7 E_i e^\dagger_i e_i  \ .
\end{equation}
Excitons also couple to vibrations whose origin may be intramolecular or related to protein motion. 
The latter can be rather delocalized conformational modes of low-frequency or collective vibrations of slightly higher energy, that may be of the same order than differences between exciton energies $E_i$ [\onlinecite{Morgan16}]. These collective vibrations are probably coupled inhomogeneously to the different BChls.  In the present paper we adopt a simpler model: vibrations are modeled by local phonon baths, reduced to one mode per site, for which the Hamiltonian reads 
\begin{equation}
H_{pn} = \sum_{i=1}^7 \omega_{0i} p^\dagger_i p_i  \ ,
\end{equation}
and their interaction with excitons is expressed as
\begin{equation}
H_{e-pn} = \sum_{i=1}^7 g_id^\dagger_i d_i (p^\dagger_i +p_i) \ .
\label{hepn}
\end{equation}
Furthermore, in the present treatment, the exciton-phonon coupling is supposed uniform: $g_i =g$. 
In the absence of this interaction, the exciton spectrum is a discrete one, through this interaction, it can acquire a continuous nature.
Finally the Hamiltonian contains also the injection and extraction terms: 
excitons are extracted at site 3, while they are injected at sites 1 and 6 [\onlinecite{Adolphs06}] :
\begin{equation}
H_T = V_c (d^\dagger_1 c +d^\dagger_6 c +hc ) +V_r (f^\dagger d_3   +hc) \ ,
\end{equation}
here $c$ annihilates an exciton in the injection reservoir called chlorosome, whereas $f^\dagger$ creates an exciton in the extraction reservoir called reaction center. 

Finally, these reservoir Hamiltonians read
\begin{equation}
H_c = E_c c^\dagger c \ \ , \ \ H_r = E_r f^\dagger f \ .
\end{equation}
In the present treatment, the influence of excitons onto their surroundings are not provided. However some feedback onto the reservoirs is considered through a 
Lorentzian broadening of $E_c$ and $E_r$ levels, as well as onto the photon levels. 
All the operators in the Hamiltonian (\ref{HamiltonienTotal}) are of bosonic character. 

\subsection{Currents}

The number of exciton per second flowing out from site $i$ can be expressed as
\begin{equation}
J_i = - \langle \frac{d N_i}{dt} \rangle = \frac{1}{i \hbar} \langle [H, N_i] \rangle \ , 
\label{Ji}
\end{equation}
where $N_i=d^\dagger_i d_i$ is the operator counting the number of exciton at site $i$. 
The commutator can be evaluated, and leads to a decomposition 
\begin{equation}
J_i = \sum_{j \neq i} J_{ij} + J^c_i +J^r_i +J^b_i \ ,
\label{Jidecomp}
\end{equation}
that can be interpreted through different events: the exciton leaving the site $i$ may flow toward the others sites $j$ ($J_{ij}$), toward the chlorosome ($J^c_i$), the reaction center ($J^r_i$), or can de-excite by emitting a photon ($J^b_i$). 

The NEGF formalism is appropriate for expressing the different currents and densities of this full open system. This formalism is also appropriate in time-dependent regimes, however 
in this paper we focus on the stationary case, in which $J_i = 0$, by the first Kirchhoff's law.
The detailed formalism for currents and densities is presented in Appendix A. 

\subsection{Parameter values}

We use the cm$^{-1}$ as energy unit, which is the one in use in this context. For definiteness 1 cm$^{-1} \simeq 0.124$ meV. 
The different input parameters needed to evaluate the various currents are listed below.\\
- The parameters $E_i$ and $V_{ij}$ are taken from Ref. [\onlinecite{Cho05}]. 
Thereafter, the local energies $E_i$ are shifted by $E_3 \simeq 12350$ cm$^{-1}$, the extraction site energy being chosen as the origin. 
Several parameter sets for $E_i$ and $V_{ij}$ are available, including  the one of Ref. [\onlinecite{Cole13}]. The latter leads almost to the same spectrum as the one adopted here. \\
- The parameter $V_b$ is chosen based on the recombination time estimation of 1 nanosecond [\onlinecite{Lambert13}], as in Ref. [\onlinecite{Pelzer14}], leading to $V_b =0.033$ cm$^{-1}$ [\onlinecite{remarque}].\\
- The phonon frequencies $\omega_{0i}= \omega_0$ are assumed to be site-independent and small compared to temperature in the so-called high temperature limit.
This assumption also implies that they are very small compared to exciton energies, such that exciton-vibration interaction will be considered as elastic. 
$\omega_0$ is combined with the exciton-phonon coupling $g$ in a widely used parameter $\lambda = g^2 / \hbar \omega_0$, known as the reorganization energy.  
In the present study, $\lambda$ will be varied to address the impact of vibration on exciton transfer.\\
- In the footsteps of Ref. [\onlinecite{Pelzer14}], the extraction parameter $V_r = 16.69$ cm$^{-1}$ is fixed by the requirement of an exciton life-time at the extraction site in the picosecond range [\onlinecite{Lambert13}], 
and the injection parameter $V_c = 0.42$ cm$^{-1}$ is chosen so as to ensure a low exciton flux. Except in the discussion of Fig.~\ref{alpha-Vc}, $V_c$ will be fixed to this value in the following calculations. \\
- The chlorosome reservoir energy, $E_c$, is chosen such that the injection corresponds to the highest exciton energy of the FMO, namely the highest eigenvalue of the $\{E_i, V_{ij} \}$ matrix. In this instance,
$E_c= 480.17$ cm$^{-1}$, while the extraction is carried out at the lowest eigenvalue: $E_r = -24.14$ cm$^{-1}$.
These reservoir levels are largely broadened in the following calculations, as previously mentioned,  by a Lorentzian parameter $\eta_c = \eta_r =105$ cm$^{-1}$. This broadening is a phenomenological way to take into account hybridization between FMO and reservoir states, inside the reservoirs themselves. A broadening affects also the photon properties as discussed in the Appendix A. \\
- NEGF also requires to fix the number of pseudoparticles in the reservoirs. The exciton number in the chlorosome and in the reaction center, respectively $n_c$ and $n_r$ will vary in the following investigations. 
They constitute the "fuel" or the driving force of the exciton transport through the FMO circuit, as such they also influence the exciton number inside the FMO complex. The latter is not an input parameter, but is determined by calculation. \\
- Finally to disadvantage exciton creation inside the FMO complex itself, compared to creation inside the chlorosome, the photon number 
inside the FMO is zero. Excitons can thus recombine, but cannot be created locally by absorption.\\
- With the hypothesis of a phonon bath at equilibrium, the phonon temperature is also a parameter emerging in the treatment of exciton-phonon interaction. It will be fixed to $k_B T = 200$ cm$^{-1} \simeq 300$ K.

\section{Results}
\subsection{ Inflow and outflow currents}

Two chromophores, respectively numbered 1 and 6 are the sites for injection, while the one numbered 3 is the extraction site. Thus, the input current is the sum of two contributions from the chlorosome, whereas the output current counts the exciton per second escaping from site 3 to the reaction center:
\begin{equation}
J_{in} = J_c = - J_{1}^c - J_{6}^c \ \ , \ \ \ J_{out} = J_r  = J_{3}^r
\label{Jinout}
\end{equation}
%ici
In the Appendices it is shown that these currents can be expressed using Greens functions and self-energies (to be defined later) through
\begin{equation}
\begin{array}{lcl}
J_c  & =& - \frac 2 \hbar \mathrm{Re} \int \frac{d \omega}{2 \pi} \mathrm{Tr} \bigl[ G^r \Sigma_c^< + G^< \Sigma^a_c \bigr] \\
J_r  & =& \frac 2 \hbar \mathrm{Re} \int \frac{d \omega}{2 \pi} \mathrm{Tr} \bigl[ G^r \Sigma_r^< + G^< \Sigma^a_r \bigr]  \ . 
\end{array}
\label{Jcr}
\end{equation}
In general, these currents are not equal due to radiative recombination events onto the FMO, and the ratio $J_{r} / J_{c}$ is a measure of the transport efficiency. 

Modeling the FMO as an open quantum system implies that the only driving force behind the exciton transport lies in the excitonic populations inside 
the reservoirs, namely $n_c$ and $n_r$.
In the work of Pelzer [\onlinecite{Pelzer14}], $n_c$ and $n_r$ were fixed to 1 and 0 respectively. By varying them, we propose 
to analyze the properties of this driving. 
Calculations over a broad range of $n_c$ and $n_r$ have revealed an exact linear dependence of the currents, that can be formulated in the following matrix notation
\begin{equation}
\begin{array}{lcl} 
J_c & = & \alpha_{cc}  n_c -\alpha_{cr} n_r \\ 
J_r & = &  \alpha_{rc}  n_c -\alpha_{rr} n_r    \ .
\end{array}
\label{linearite}
\end{equation}
The signs have been chosen such that all the matrix elements in the previous equations are positive. By the Kirchhoff's law, the total radiative current is simply the difference between input and output currents: $J_{rad} = J_{c} -J_r = \sum_i J_i^b$. Thus $J_{rad}$ is also an affine function of the reservoir populations. 
The $\alpha$-matrix elements can be expressed in terms of retarded ($G^r$) and advanced ($G^a$) FMO Green's functions (GF), and various self-energy terms, which are detailed in the Appendices, where the following  expressions are derived:
\begin{equation}
\begin{array}{lcl}
\alpha_{cc} & = & - \frac 2 \hbar \mathrm{Re} \int \frac{d \omega}{2 \pi} \mathrm{Tr} \bigl[  G^r \tilde{\Sigma}^<_c  + G^r \bigl( \tilde{\Sigma}^<_c + X_c \bigr) G^a \Sigma^a_c    \bigr] \\
\alpha_{cr} & = &  \frac 2 \hbar \mathrm{Re} \int \frac{d \omega}{2 \pi} \mathrm{Tr} \bigl[ G^r \bigl( \tilde{\Sigma}^<_r + X_r \bigr) G^a \Sigma^a_c    \bigr] \\
\alpha_{cr} & = &  \frac 2 \hbar \mathrm{Re} \int \frac{d \omega}{2 \pi} \mathrm{Tr} \bigl[  G^r \bigl( \tilde{\Sigma}^<_c + X_c \bigr) G^a \Sigma^a_r    \bigr]  \\
\alpha_{rr} & = & - \frac 2 \hbar \mathrm{Re} \int \frac{d \omega}{2 \pi} \mathrm{Tr} \bigl[  G^r \tilde{\Sigma}^<_r  + G^r\bigl( \tilde{\Sigma}^<_c + X_c \bigr) G^a \Sigma^a_r    \bigr] \ . 
\end{array}
\label{alphacoeff}
\end{equation} 
We have defined the tilde notation by dividing the self-energy by density: $\tilde{\Sigma}_\alpha^<  = \Sigma_\alpha^< /  n_\alpha$, for $\alpha=r,c$ (see Appendix C).
As shown in the Appendix C, the linear dependence relies on the fact that the injection and extraction reservoirs accommodate only one level, and
on the hypothesis of a phonon bath at equilibrium, at a temperature high enough such that $k_B T \gg \hbar \omega_{0i}$.
However the assumption of an elastic exciton-phonon interaction, made in the present study, is not a necessary ingredient for linearity. 
By symmetry we expect $\alpha_{cr} = \alpha_{rc}$, as established in Appendix C. 

\begin{figure}
\includegraphics[width=0.5\textwidth]{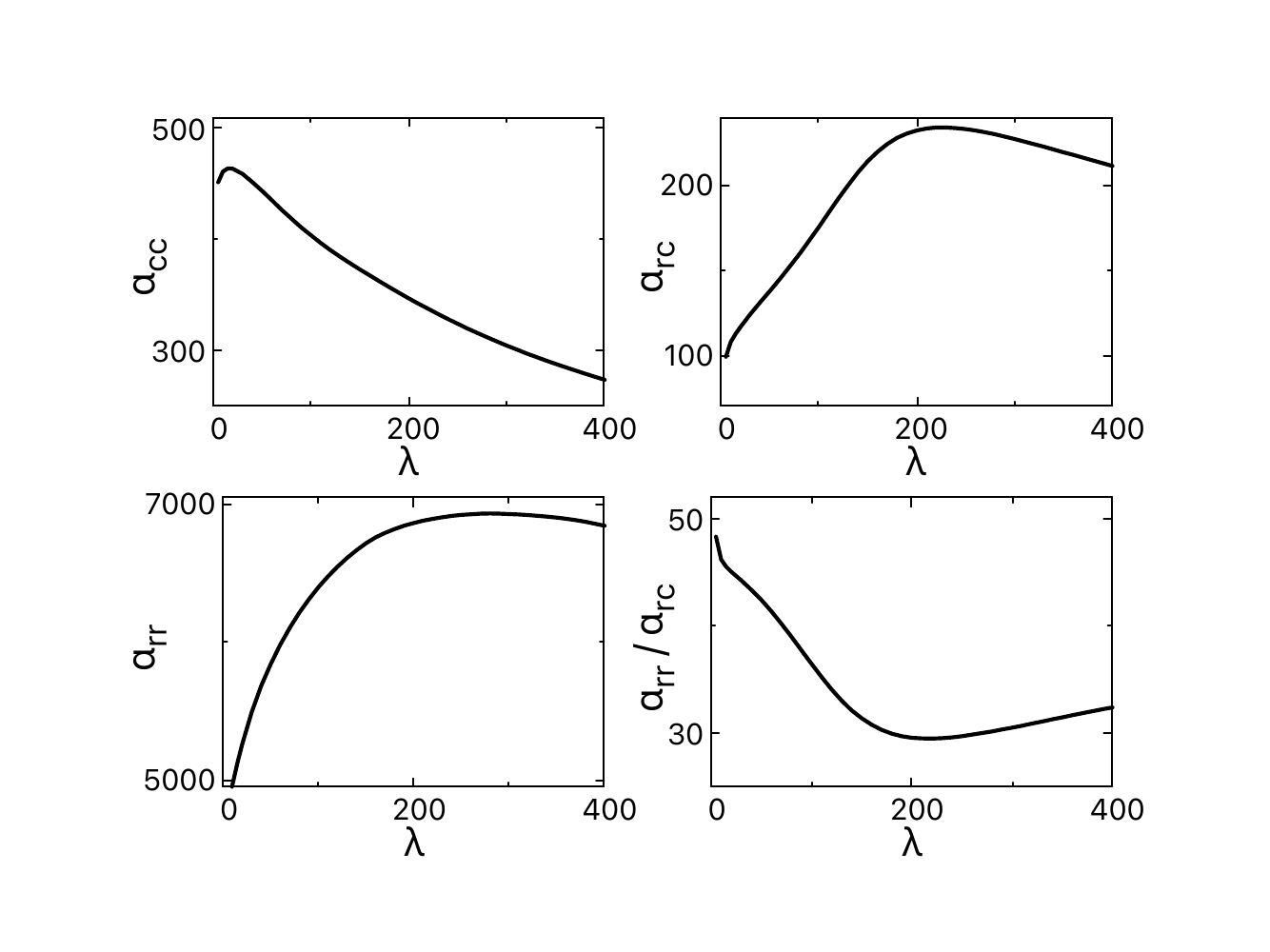}
\caption{Coefficients $\alpha_{cc}$, $\alpha_{rc}$, $\alpha_{rr}$, and $\alpha_{rr}/\alpha_{rc}$, as defined in Eq.~(\ref{linearite}), expressed in MHz, a a function of the reorganization energy $\lambda$ in cm$^{-1}$.}
\label{alpha}
\end{figure}

The different coefficients 
$\alpha_{cc}$, $\alpha_{rc}$ and $\alpha_{rr}$ are plotted in Fig.~\ref{alpha} as a function of the reorganization energy $\lambda$, which measures the strength of the exciton-vibration interaction.
The three coefficients first raise, then decrease, as $\lambda$ increases. This non-monotonic behavior for $\alpha_{cc}$ and $\alpha_{rc}$ was also observed in 
Pelzer's work [\onlinecite{Pelzer14}] and explained in terms of  broadening of the exciton spectral function due to exciton-vibration interaction. This interaction is 
favorable to transport at low to moderate interaction, before becoming detrimental at higher coupling value, due to coherence loss.  The maximum for $\alpha_{cc}$ is attained for a rather small and sharp value 
of $\lambda$ close to $\lambda = 20$ cm$^{-1}$, whereas the growth of the other two goes on until a larger value of $\lambda$ lying in between 200 to 300 cm$^{-1}$. 
The precise location of these extrema, as well as their values, are closely related to injection and extraction levels, as revealed by 
numerical calculations.  
It should be noted that $\alpha_{cc}$ and $\alpha_{rc}$ can even be raised for an injection at the second eigenmode, namely $E_c=408.2$ cm$^{-1}$, due to a bigger 
overlap between the
entry points and the second eigenvector, than between the entry sites and the first eigenvector. For the same reason of larger overlap between the exit point and the lowest eigenmode, the extraction energy $E_r$ adopted
here, is optimal from the efficiency point of view. 
In the FMO complex, the order of magnitude of the reorganization energy would be about 40 cm$^{-1}$ [\onlinecite{kramer14,Fokas17}],  but values up to hundreds cm$^{-1}$ have been reported [\onlinecite{Jancovic20}].

One sometimes reads - for example on the wikipedia entry for "Fenna-Matthews-Olson complex" entry - that the onsite BChl energies  $\{E_i\}$ determine the energy 
flow from chlorosome to reaction center, in some funnel or cascade picture. 
The present study is not the first one to challenge this idea [\onlinecite{Brixner05}], however it sheds a different light on this idea. 
As widely acknowledged, the $\{E_i\}$ distribution defines the extent of the FMO energy spectrum. 
The vibrations, even treated in an elastic approximation, broaden the otherwise narrow exciton spectral peaks and settle 
overlaps that entitle the exciton transfer. However it is the reservoir exciton populations that are the key drivers of the transfer process. 
In other words, from Eq. (\ref{linearite}), we observe that the exciton flow travels from the reaction center toward the chlorosome for $n_c=0$ and $n_r=1$; and the 
current which in turn becomes the output one $J_c = - \alpha_{rc}$
 is the same, just reversed, than the current flowing out from the FMO for $n_c=1$, $n_r=0$.
Due to the asymmetry $\alpha_{cc} \neq \alpha_{cr}$,
nor is it the density gradient, of which $(n_c -n_r)$ would be the witness, that drives the transport. 
As shown in the next paragraph, a small $n_r$ value can change the output current direction. So the sometimes raised energy gradient does not seem so relevant. 

In addition to not being monotonous functions of the reorganization energy, the $\alpha$ matrix elements are not all of the same order of magnitude. 
The value of the ratio $\alpha_{rr}/\alpha_{rc}$, also plotted in Fig.~\ref{alpha}, is linked to the present reservoir coupling ratio $\frac{V_r}{V_c}$ close to 40.

\subsection{Redesigning the FMO complex}
The symmetry of the matrix $ V_{ij} $ reduces the coupling number to 21 [\onlinecite{number}]. Their absolute values taken from Ref. [\onlinecite{Cho05}] lie within the range $[106-1 ]$ cm$^{-1}$. 
In an attempt to acquire 
 a deeper understanding of the $\alpha$-matrix coefficients and their relation to the FMO architecture, we first disregard the smallest hopping elements of the $V_{ij}$ 
matrix, by removing the terms whose absolute values are smaller than 20 cm$^{-1}$. Seven coupling terms remain, they are shown in solid lines in 
Fig.~\ref{FMObase}, and  
reveal two main parallel  transport paths, one of them including a loop. We call this structure the {\it stiff} one. 
As can be expected, the  designed pathways bear some resemblance to those that were selected as the main relaxation paths in the FMO [\onlinecite{delRey13,Adolphs06,Brixner05,Ishizaki09}].

\begin{figure}
\includegraphics[width=0.5\textwidth]{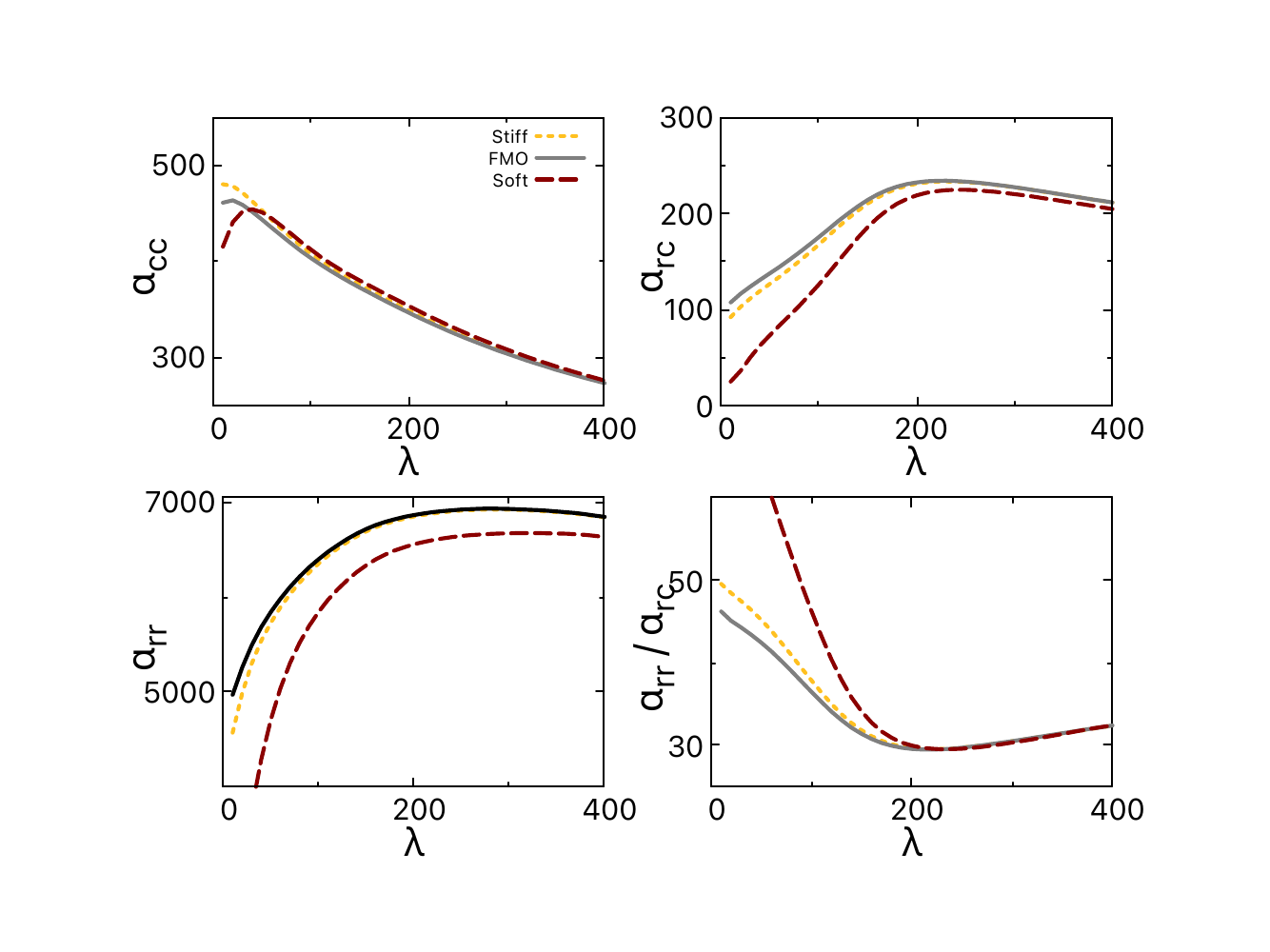}
\caption{$\alpha$-matrix coefficients for different FMO architectures: the normal one (FMO, same as in Fig.~\ref{alpha}), and the {\it stiff} and {\it soft} ones. They are expressed in MHz, as a function of the reorganization energy $\lambda$ in cm$^{-1}$. See text for further details.}
\label{CurrentFull-and-Cut}
\end{figure}

The impact of this truncation on the $\alpha$-matrix coefficients is shown in Fig.~\ref{CurrentFull-and-Cut}. 
In this architecture calculation, the reservoirs energy remain respectively the highest and lowest eigenvalues of the {\it normal} FMO.
For $\lambda \le 100$ cm$^{-1}$, the 
simplified stiff structure overestimates $\alpha_{cc}$ and underestimates the other two by a few percent only. For a higher value of $\lambda$, the values merge, revealing 
the absence of influence of the smallest $V_{ij}$ terms on external currents in this exciton-vibration coupling range.  
In the same Fig.~\ref{CurrentFull-and-Cut}, the $\alpha$ matrix elements corresponding to the opposite case are also overlaid: an FMO architecture for which the major 
couplings have been withdrawn, keeping only those that were previously neglected. We call this network the {\it soft} FMO. 
Again, the reservoirs energy remain unchanged.
The corresponding $\alpha$ coefficients are further 
away from the real structure, but surprisingly there is no collapse of the current.

This can be understood by the following arguments: first of all, all the three considered structures still involve all the seven BChls. Furthermore, diagonalizing the corresponding 
$\{E_i, V_{ij}\}$ matrix leads to eigenvalues, which are dominated by the $\{ E_i \}$, and which are very close for normal and stiff FMO structures. 
This predominance of $\{E_i\}$ over $\{ V_{ij} \}$ explains why the optical spectrum is extremely sensitive to the choice of on-site energies, as underlined in Ref. [\onlinecite{Milder10}].
The eigenvalues for the soft FMO architecture constitute a narrowed spectrum compared to the one pertaining to the real FMO: this  explains that the $\alpha$ coefficients are reduced
in that case.

This comparative study between normal, {\it soft} and {\it stiff} architectures underlines the robustness of the FMO complex: the $V_{ij}$, which are very sensitive to the 
mutual orientation of the BChls, may change in a noisy environnement, without affecting too much the external currents as we have shown.
This is in line with the conclusions of other works - see for example Refs. [\onlinecite{Worster19,Harush21}] - that have demonstrated that the complex structure plays only a minor role in the determination of the currents. 

\subsection{Toward a new regulation mechanism?}

We now address the question of the influence of the exciton population inside the reaction center.
In Pelzer's work [\onlinecite{Pelzer14}], the exciton extraction from the reaction center toward photosynthetic
chemical reactions is 
supposed to be so efficient and rapid, that excitons are rapidly consumed inside the reaction center and $n_r=0$ is assumed all along. 
In the following, we depart from this very special case.

In addition, we have not drawn all the consequences from Eq. ~(\ref{linearite}), and especially from the difference in order of magnitude between the coefficients:
 $\alpha_{rr} \gg \alpha_{rc}, \alpha_{cc}$.
Thus, for an exciton number onside the reaction center verifying $ n_ r \ge n_c \frac{\alpha_{rc}}{\alpha_{rr}}$, the output current can be reversed, well before the input 
one is reversed too: 
to fix the ideas, for $\lambda= 100$ cm$^{-1}$, a small exciton number $n_r \ge 0.028$  for $n_c =1$ is enough to reverse the output flow, whereas passing from $n_r=0$ 
to $n_r=0.028$ only reduces $J_{c}$ by less than 1.5\%.  

\begin{figure}
\includegraphics[width=0.5\textwidth]{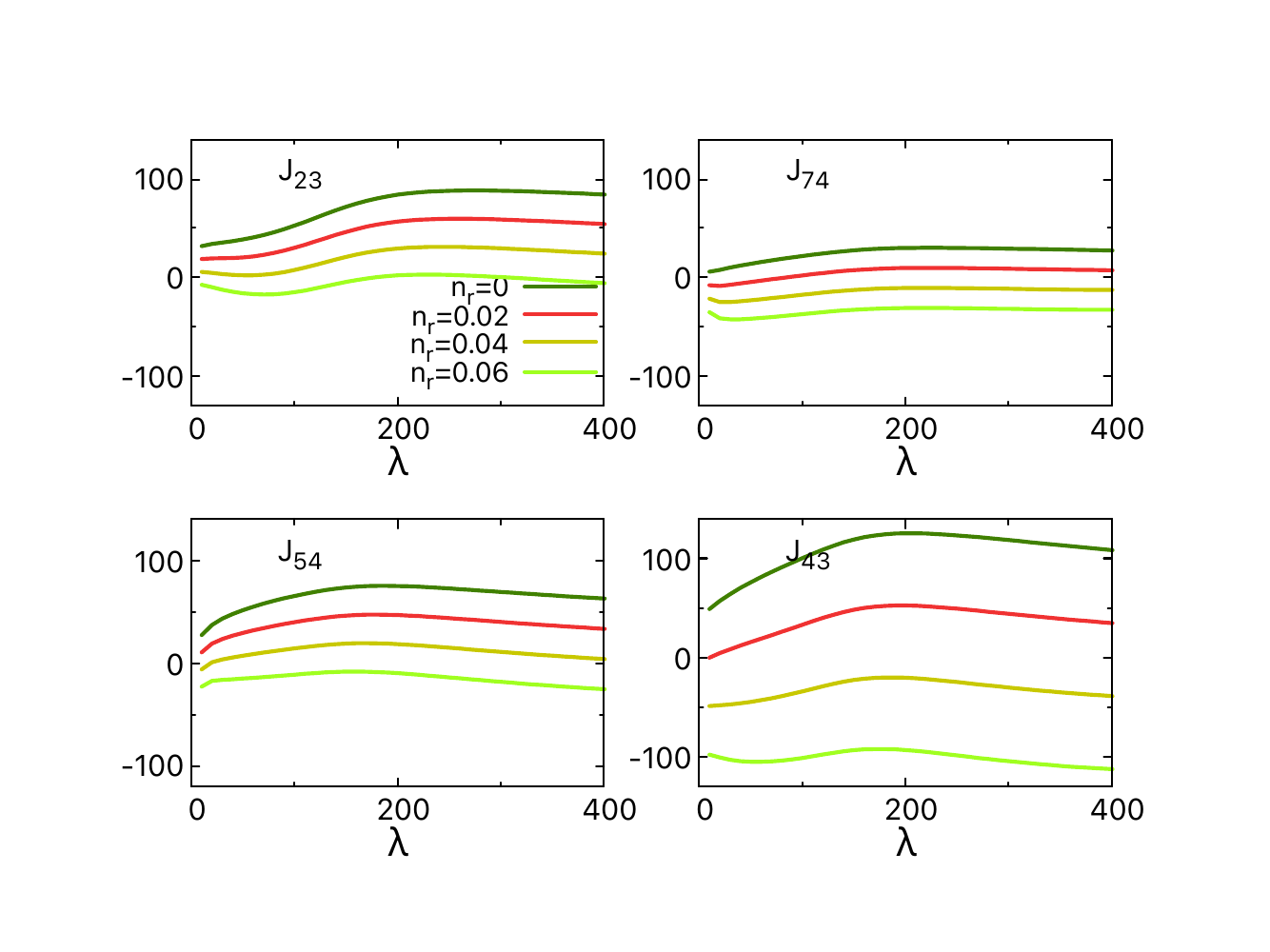}
\caption{ Main inter-BChl FMO currents as a function of $\lambda$, for four different $n_r$ values, and $n_c=1$.
}
\label{MainLocalCurrents}
\end{figure}

In addition to the injection and extraction currents, NEGF calculations enable also to evaluate the inter-BChl currents, previously defined as $J_{ij}$.
From now on, the calculations will be done for the previously called {\it normal} FMO, that is, standard FMO. 
The main currents correspond to the highest $|V_{ij}|$ values, and we focus on them, namely
$J_{65}$, $J_{67}$, $J_{54}$, $J_{74}$, $J_{43}$, $J_{12}$, and $J_{23}$, for different values of $n_r$, but a fixed $n_c$ value. 
In the present convention $J_{ij}$ is positive when excitons leave site $i$ for site $j$. 
The local currents, as a function of $\lambda$ are displayed in Fig.~\ref{MainLocalCurrents} for the four more important ones, while the other three, in addition to the 
output current, are displayed in the Appendix D. They are all plotted for $n_r=0$ and for three low $n_r$ values: namely 0.02, 0.04 and 0.06, while for all of 
them $n_c= 1$. 
For an empty reaction center ($n_r=0$), all the previous local currents are positive, but they are not of the same amplitude: the branch with the loop is more loaded
than the simpler one:
indeed $J_{65} +J_{67} > J_{12}$, and $J_{43} > J_{23}$.  
For $n_r = 0.02$, a change of sign appears in $J_{74}$ at low $\lambda$, whereas for $n_r=0.04$, it is negative for any $\lambda$ 
value. In this regime, $J_{67}$ remains positive, without violating Kirchhoff's law, mainly due to high radiative recombination at site $7$.
Overall, $J_{43}$, and as a consequence $J_r$, are the most sensitive to a small exciton population on the reaction center.
Indeed,  for $\lambda = 100$ cm$^{-1}$, the output current is now also negative: the small value of $n_r=0.04$ reverses the output current, meanwhile, 
the input one is barely reduced.
This means that all the exciton energy is dissipated on the FMO itself by radiative recombination. 

The observed sensitivity of the output flow to the value of $n_r$ could be an avenue for a protection mechanism of the reaction center, which, to our knowledge has not yet been
 mentioned.
Numerous studies point to the existence of mechanisms for protecting the reaction center from overheating that are not yet fully elucidated
[\onlinecite{Orf16,Magdaong18,Higgins21,Klinger23}].
The hypothesis put forward in this study is as follows: 
beyond a certain exciton density at the reaction center - albeit much lower than the density on the chlorosome or on the FMO itself - the output current reverses, 
despite an ongoing injection of exciton from the chlorosome. The number of excitons at the reaction center then decreases, as a result of increased radiative recombination inside the FMO itself. Furthermore, this possible feedback mechanism to prevent overheating could be quite reactive, since the related transfer time scale is of the order of a picosecond.

\begin{figure}
\includegraphics[width=0.45\textwidth]{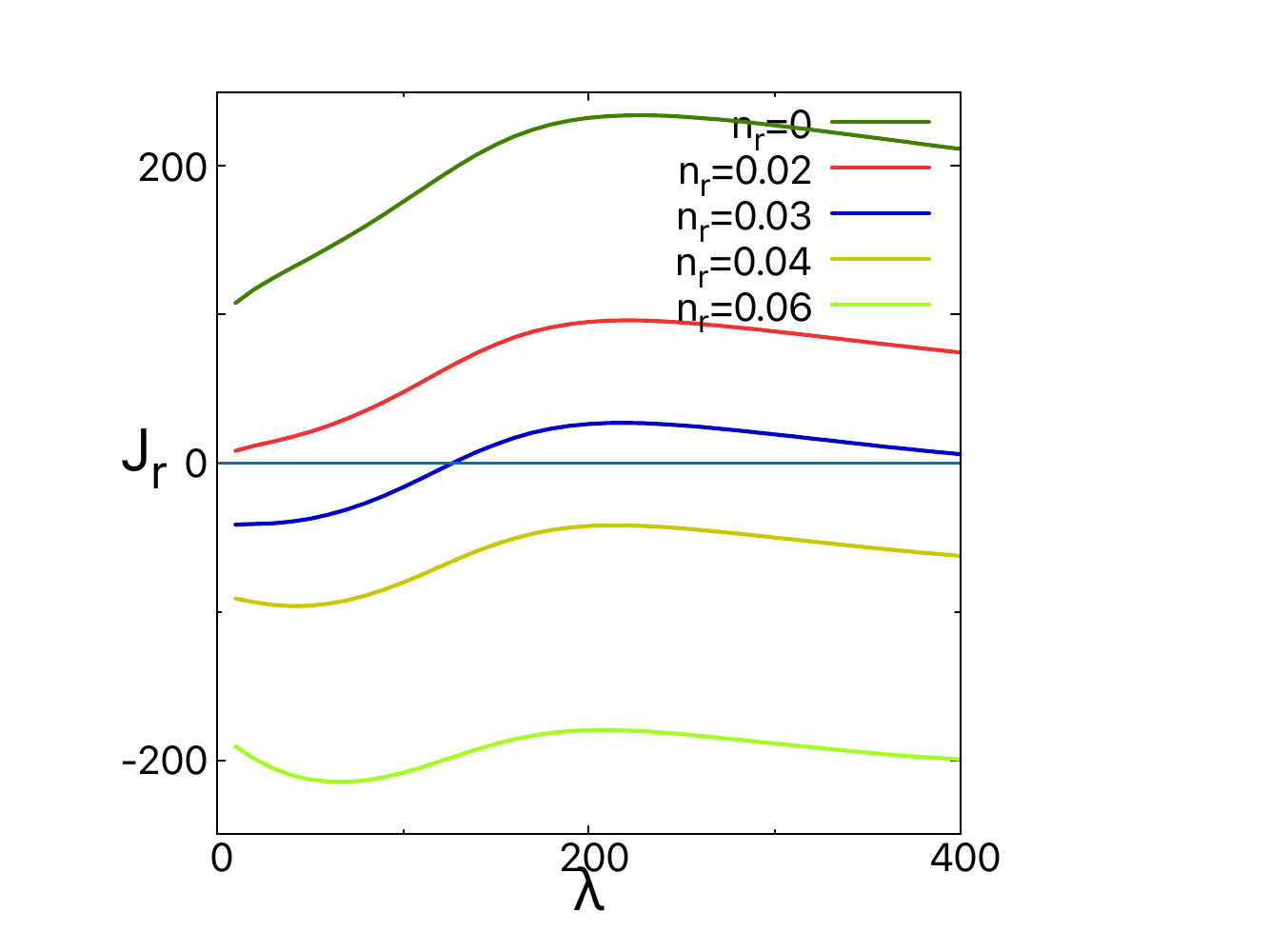}
\caption{ Output FMO currents $J_r$, in MHz, as a function of $\lambda$ in cm$^{-1}$,
for different $n_r$ values.  An horizontal line corresponding to zero current guides the eye.
}
\label{Jr-2-nr}
\end{figure}

An alternative regulation mechanism can be also invoked, taking advantage of the sensitivity of the output current to the value of the reorganization energy.
Indeed it has been suggested that to alleviate overheating of the reaction center, a conformal change may occur [\onlinecite{{scholes11}}].
It is certain that such a molecular reorganization would result in a modification of the $\lambda$ value. We have demonstrated that output currents can decrease or even reverse as $\lambda$ varies, as shown quantitatively in Fig.~\ref{Jr-2-nr}. The non-monotonic function $J_r(\lambda)$ may entitle a variation of $\lambda$ in both directions. 

The above-mentioned protection mechanisms rely on the possibility to reverse the output current for a low exciton density at the reaction center. 
From the expression of the output current: $J_r =\alpha_{rc} n_c -\alpha_{rr} n_r$,  and reservoir densities satisfying $n_r \ll n_c$, reversing the output flow requires
a large $\alpha_{rr} / \alpha_{rc}$ ratio, which is related to $V_r / V_c$. 
We thus address the question of robustness of the inequality $\alpha_{rr} / \alpha_{cr} \gg 1$. 
In Pelzer's modeling, and in the present calculations, the value of the parameter $V_c$ that couples FMO complex and reaction center was chosen on the basis of a
low exciton flux through the FMO. 
 The question is whether the range of values for this parameter is wide or not.
 \begin{figure}
\includegraphics[width=0.5\textwidth]{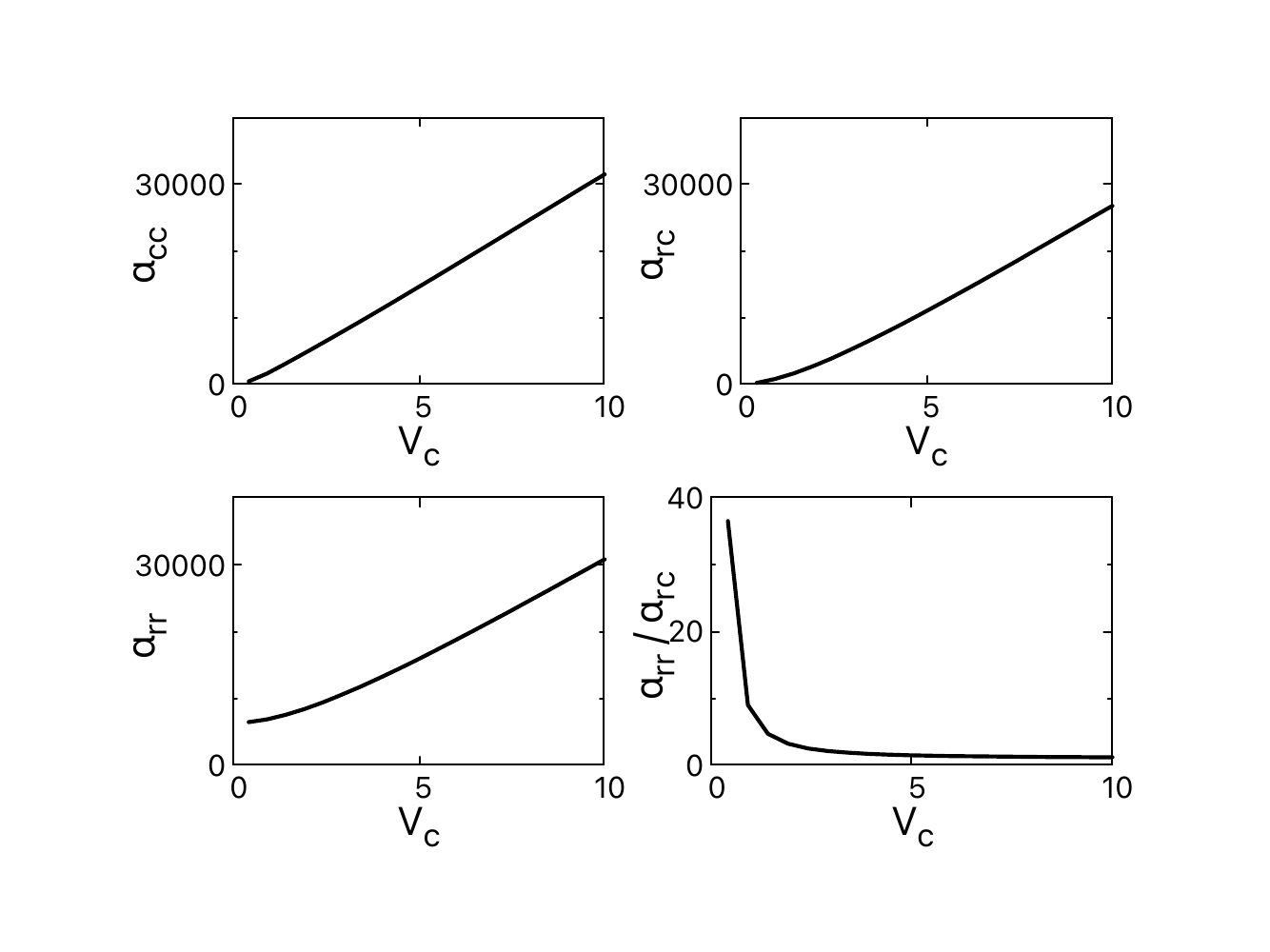}
\caption{$\alpha$-matrix coefficients  in MHz for $\lambda=100$ cm$^{-1}$ as a function of $V_c$, the coupling between FMO and chlorosome. The 
coupling parameter $V_r$ between FMO and reaction center, is fixed as previously $V_r=16.69$ cm$^{-1}$. 
}
\label{alpha-Vc}
\end{figure}
Fig.~\ref{alpha-Vc} displays the different coefficients of the $\alpha$-matrix as a function of $V_c$, for fixed $V_r=16.69 $ cm$^{-1}$ and $\lambda=100$ cm$^{-1}$.
For $V_c \ge 5$ cm$^{-1}$, 
the three coefficients $\alpha_{cc}$, $\alpha_{rc}$ and $\alpha_{rr}$ are of the same order of magnitude, and as a consequence
$\alpha_{rr} / \alpha_{rc} \rightarrow 1$.
This figure reveals that, for $\lambda=100$ cm$^{-1}$, $V_c \lessapprox 1$ cm$^{-1}$, or equivalently $V_r / V_c \gtrapprox 20$, is required to guarantee  a $\alpha_{rr} / \alpha_{rc}$ ratio greater than 10, which is necessary for the proper functioning of the proposed regulation mechanisms.

\subsection{Density profiles}

\begin{figure}
\includegraphics[width=0.5\textwidth]{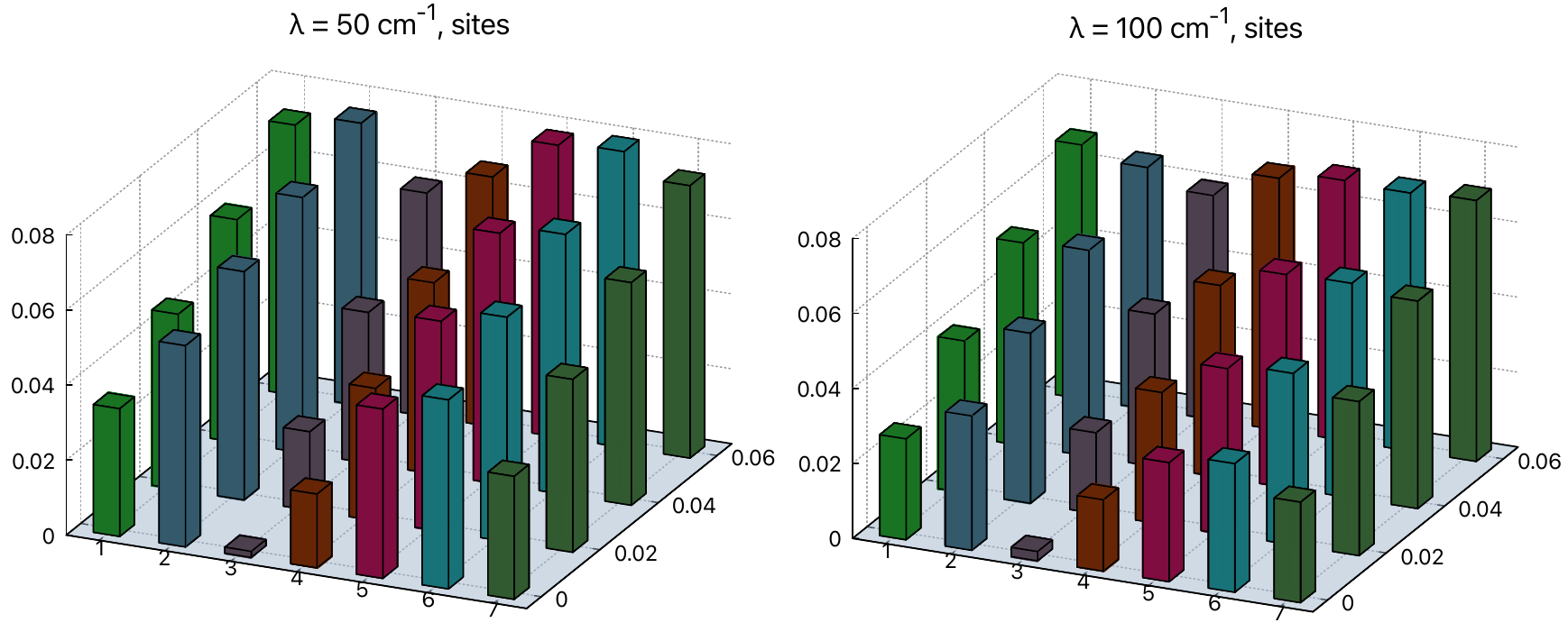}
\caption{ Site populations as a function of the site number (i=1, ..7) for $n_r=0, 0.02, 0.04$ and 0.06, for two different values of the reorganization energy. Throughout, $n_c=1$.
}
\label{SitePopulations}
\end{figure}

Finally, it is instructive to look at the exciton density along the FMO sites as $\lambda$ or $n_r$ vary. 
As mentioned in Appendix A, local density and local recombination current are proportional in the present modeling. 
The density profile is shown in Fig.~\ref{SitePopulations} for $\lambda=50$ cm$^{-1}$ and $\lambda=100$ cm$^{-1}$, and
for different $n_r$ densities at the reaction center. $n_c=1$ is assumed throughout this study. 
It reveals a very efficient exciton evacuation at site 3, for an empty reaction center: $n_3$ is 
lower than the other $n_i$, at least by an order of magnitude. 
Increasing $n_r$ raises the total exciton density and also tends to homogenize it on the whole FMO. 
Overall, $n_3$, although the lowest local density, is larger than $n_r$.
Spreading of the exciton population is also the result of a exciton-vibration coupling increase as can be seen from Fig.~\ref{SitePopulations}, comparing $\lambda=50$ and 100 cm$^{-1}$. A slight decrease of the total population accompanies it. 
On this figure, the total density at most reaches 0.5 for $n_r=0.06$ and $\lambda=50$ cm$^{-1}$.

\begin{figure}
\includegraphics[width=0.5\textwidth]{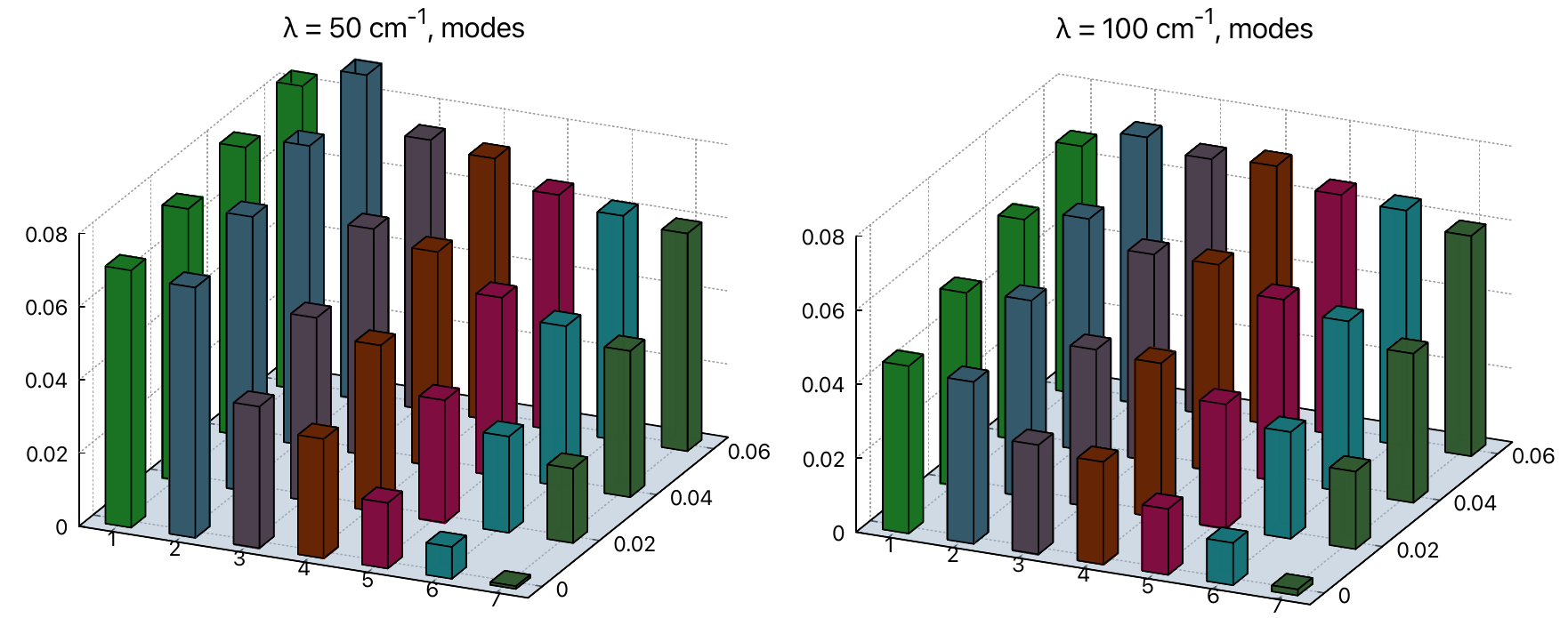}
\caption{Mode populations as a function of the mode number (i=1, ..7) and $n_r$, for two different values of the reorganization energy.  The parameters are the same as in Fig.~\ref{SitePopulations}. 
}
\label{ModePopulations}
\end{figure}

One can discuss the exciton densities from the perspective of eigenmodes instead of BChl sites.
Fig.~\ref{ModePopulations} presents the local densities distributed on the different FMO eigenmodes, which are indexed in order of decreasing eigenvalues.
The observed decrease of the mode populations as the mode index raises for $n_r=0$, can be understood, from the working condition used: 
injection from chlorosome is done at $E_c$ equal to the highest eigenvalue, while extraction to reaction center $E_r$ is done at the lowest one, as stated in the parameter values paragraph. 
The mode populations as well as the site ones, are spread by $\lambda$, as expected, due to energy broadening. 
A less predictable behavior is the fact that as $n_r$ raises, the maximum mode population shifts towards smaller energy. This is reinforced for higher $\lambda$ values.

\section{Conclusion}

In the present paper, we take a fresh look at local exciton transport in the FMO photosynthetic complex by
using NEGF, a technique suitable to open quantum systems.
We have re-examined the widespread idea of energy cascade or funnel, and proposed the following off-center point of view: the important fact for efficient FMO
operation is injection at an energy matching the eigenvalue corresponding to the eigenvector with great overlap with the BChls identified as entry points, and symmetrically, 
extraction at the eigenvalue corresponding  to the eigenvector with greatest overlap with the BChl identified as exit point. 
We have also found that broadening of these eigenstates by interaction with vibrations is also an essential point for the smooth operation of the FMO complex. 

We have established that the fuel of exciton transport through the complex is the exciton density on the reservoirs (chlorosome and reaction center), 
the input and output currents being affine functions, in our modeling, of the respective excitonic densities $n_c$ and $n_r$.

The asymmetry between the affine coefficients opens up a possibility of two protection mechanisms, one of which has not yet been mentioned. 
The latter can be summarized as follows: at low reaction center filling, the flow between FMO and reaction center is reversed. 
This releases energy through recombination.
This protection mechanism might be also desirable for artificial nanodevices. If not, it can be removed by a proper choice of the coupling between the central system 
and its reservoirs.

Although we do not expect these conclusions to be fully invalidated,
it is important to test the robustness of the current results within the framework of a more realistic model for exciton-vibration interaction. Indeed, the present calculations make 
essentially two assumptions: high temperature compared to typical vibration frequencies, and elastic exciton-vibration interaction, whereas room temperature 
 is of the same order of magnitude than typical intramolecular vibrational frequencies [\onlinecite{Wendling00,Lee16}].
This improvement constitutes a numerical challenge, which is currently being worked on.

\section*{Acknowledgments} 
We thank G. Ayala, M. Barbatti, M. Bescond, A. W. Chin, and J. Toldo  for valuable discussions.

\section*{Appendix A: Green's functions for currents}

We develop the formalism of NEGF in this Appendix. 
Numerous textbooks and pedagogical papers  [\onlinecite{HaugJauho,Maciejko07,Jauho06,Wang14}],  can be found, they usually concern fermions rather than bosons.
Even if we are interested in the present paper in the stationary regime, NEGF calculation developments need time dependencies.

The goal of the present calculation is to obtain the various terms involved in Eq. (\ref{Ji}), where $N_i$ is the operator counting the number of exciton at site $i$ in 
Heisenberg representation.
Evaluation of the commutator leads into a decomposition for $J_i$ as written in Eq. (\ref{Jidecomp}), where the current between sites $i$ and $j$ can be expressed as follows
\begin{equation}
J_{ij}(t)=  \frac 2 \hbar \mathrm{Re}\Bigl[ V_{ji} G^<_{ij} (t,t)  \Bigr] \ ,
\label{Jij}
\end{equation}
in terms of a lesser Green function  (GF) which  is defined by
\begin{equation}
 G^<_{ij}(t,t') = - i \langle d_j^\dagger (t') d_i (t)\rangle \ .
 \label{GF1}
\end{equation}
In Eq. (\ref{Jij}), the GF is an equal time version of Eq. (\ref{GF1}).
Eq. (\ref{Jidecomp}) also includes currents between the site $i$ and one of the reservoirs, namely chlorosome (c) or reaction center (r), and finally between the site $i$ and the local photon-bath (bi). We designate all these terms by the generic $\alpha$, and their evaluation through the commutator lead to
\begin{equation}
J_i^\alpha (t) = \frac 2 \hbar \mathrm{Re}\Bigl[ V_{\alpha i} G^<_{i\alpha} (t,t)  \Bigr] \ ,
\label{jialpha}
\end{equation}
where,  
for $\alpha=r$, $V_{r i} = V_r \delta_{i3}$ and $G^<_{ir} (t,t') = - i \langle f^\dagger (t') d_i (t)\rangle$; 
for $\alpha=c$, $V_{c i} = V_c (\delta_{i1} +\delta_{i6})$ and $G^<_{ic} (t,t') = - i \langle c^\dagger (t') d_i (t)\rangle$,
finally for $\alpha = bi$, $V_{bi i} = V_b$ and $G^<_{ibi} (t,t') = - i \langle e_i^\dagger (t') d_i (t)\rangle$. 

\subsection{Hybrid Green functions}

We treat separately what we call hybrid GF, that is $G^<_{i\alpha} $ for $\alpha=r,c$ or $bi$, and postpone $G^<_{ij}$ for a while.
To fix the notations, let us focus on $\alpha = c$. We need to define another GF named time-ordered one, as follows
\begin{equation}
G^t_{ic} (t,t')= -i \theta(t-t') \langle d_i(t) c^\dagger(t') \rangle - i \theta(t'-t) \langle c^\dagger(t') d_i(t)  \rangle
\end{equation}
which, being an equilibrium Green's function depends only on $t-t'$. 
Henceforth we adopt $\hbar =1$. 
Using $[H, c^\dagger ] = E_c c^\dagger+\sum_j V_{cj} d^\dagger_j$, we obtain
\begin{equation}
-i  \frac{\partial}{\partial t'}G^t_{ic}(t,t') = E_c G^t_{ic}(t,t') +\sum_j V_{cj} G^t_{ij} (t,t')  \ , 
\end{equation}
whose solution can be written
\begin{equation}
G^t_{ic}(t,t') =  \sum_j V_{cj} \int dt'' G^t_{ij}(t,t'') g^t_{c} (t'',t') \ ,
\label{perturbeq}
\end{equation}
with 
$g^t_{c} (t'',t') = g^t_{c} (t''-t')$, that can simply be related to the chlorosome population $n_c$, just using 
\begin{equation}
\begin{array}{lcl} 
g^t_c(t) & =  & - i \theta(t) \langle c(t) c^\dagger \rangle - i \theta(-t) \langle c^\dagger  c(t)  \rangle \\
 & = & -i \bigl( n_c  +\theta(t)\bigr)e^{-i E_ct} \ ,
\end{array}
\end{equation} 
$c^\dagger$ means here $c^\dagger(t=0)$. 
According to the rules of non-equilibrium calculations [\onlinecite{HaugJauho}], relying on the previous Eq. (\ref{perturbeq}),
we can now evaluate the function we are looking for:
\begin{equation}
\begin{array}{lcl} 
G^<_{ic} (t-t') = && \sum_j V_{cj} \int dt'' \bigl[ G^r_{ij}(t-t'') g^<_{c}(t''-t') + \\
&&  G^<_{ij}(t-t'') g^a_{c}(t''-t') \bigr]  \ ,
\end{array}
\end{equation}
that gives, in term of Fourier transform  (FT)
\begin{equation}
G^<_{ic} (\omega) = \sum_j V_{cj} \bigl[ G^r_{ij}(\omega) g^<_{c}(\omega)+ G^<_{ij}(\omega) g^a_{c}(\omega)\bigr] \ .
\label{glessic}
\end{equation}
New functions have appeared, namely $G^r_{ij}(t)$, $g^<_c(t)$ and $g^a_c(t)$, defined respectively by
\begin{equation}
\begin{array}{lcl} 
G^r_{ij}(t)  & = & - i \theta(t) \Bigl(\langle d_i(t) d^\dagger_j \rangle - \langle d^\dagger_j  d_i(t) \rangle \Bigr)\\
g^<_c(t) & = & - i \langle  c^\dagger c(t) \rangle  \\
g^a_c(t) & = &  i \theta(-t) \Bigl(\langle c(t) c^\dagger \rangle - \langle c^\dagger  c(t) \rangle \Bigr) \ ,
\end{array}
\end{equation}
where superscript indicates retarded (r), lesser ($<$), or advanced (a) GF. The last two equations concern the disconnected chlorosome, 
their respective FT can be evaluated and give
$g^<_{c}(\omega) = 2 i  n_c \mathrm{Im}( g^r_c(\omega))$ 
and 
$g^{r,a}_{c} (\omega)=1 / \bigl( \omega-E_c \pm i \eta_c \bigr) $, with $\eta_c$ a Lorentzian broadening coefficient affecting the exciton level inside the chromophore and already introduced in the main text.
The equation needed to express $G^r_{ij}(t)$ or more precisely its FT will be discussed in the next paragraph.

The preceding calculations can be replicated to evaluate $G^<_{i r} $ and $G^<_{i bi} $. This leads to 
\begin{equation}
\begin{array}{lcl} 
G^<_{ir} (\omega) & = & \sum_j V_{rj} \bigl[ G^r_{ij}(\omega) g^<_{r}(\omega)+ G^<_{ij}(\omega) g^a_{r}(\omega)\bigr]  \\
G^<_{ibi} (\omega) & = & \sum_j V_{b} \bigl[ G^r_{ij}(\omega) g^<_{bi}(\omega)+ G^<_{ij}(\omega) g^a_{bi}(\omega)\bigr] \ ,
\label{GFlesshyb}
\end{array}
\end{equation}
in which lesser GF for exciton on reaction center or photon in the bath read 
$g^<_{r}(\omega) = 2 i  n_r \mathrm{Im}(g^r_r(\omega))$ and  $g^<_{bi}(\omega) = 2 i   n_{bi} \mathrm{Im}(g^r_{bi}(\omega))$, while retarded and advanced ones read
$g^{r,a}_{r} (\omega)=1 / \bigl( \omega-E_r \pm i \eta_r \bigr) $ and $g^{r,a}_{bi} (\omega)=1 / \bigl( \omega-E_{i} \pm i \eta_b \bigr) $. $n_r$ is the exciton population at reaction center, already met in the main text. 
We have defined $n_{bi}$ as the photon number at site $i$: in the present calculations it will be set to zero to avoid exciton generation onto the FMO.  $\eta_r$ broadens the reaction center level, $\eta_b$ that of the photon. 
In Pelzer's model and in our own, $\eta_b$ is large enough to ensure constant purely imaginary $g^{r}_{bi} (\omega) = - i \pi$ green functions. In that case it can be shown that the local 
recombination current $J_i^{bi}$ is proportional to the local exciton density $n_i$ defined later. 

\subsection{FMO Green functions}

\subsubsection{$G^r_{ij}$ }

As seen above, in Eqs. (\ref{glessic}) and (\ref{GFlesshyb}) and as will be even more apparent later, we need also to evaluate $G^r_{ij}(\omega)$, in addition to $G^<_{ij}(\omega)$. 
NEGF sets out the requirements to evaluate these frequency-dependent lesser and retarded GF. 
The retarded ones are defined as 
\begin{equation}
G^r_{ij}(t) = -i \theta(t) \bigl(\langle  d_i(t) d^\dagger_j \rangle -\langle  d^\dagger_j d_i(t)  \rangle \bigr)  \ ,
\end{equation}
from which the time-derivative can be evaluated and gives
\begin{equation}
\begin{array}{lcl}
\frac{\partial}{\partial t} G^r_{ij}(t) &&= -i \delta(t) \delta_{ij} - i E_i G^r_{ij}(t) \\
& & -i \sum_l V_{il} G^r_{lj}(t)  -i g_i G^r_{pni,j}(t) -i V_{bii} G^r_{bij} (t)   \\
& & -i  V_{ci} G^r_{cj}(t) -i  V_{ri} G^r_{rj}(t) \ .
\label{deriv}
\end{array}
\end{equation}
Again, new GF have been added, defined by 
\begin{equation}
\begin{array}{lcl}
G^r_{bij}(t) &=& -i \theta(t) \Bigl( \langle e_i(t) d_j^\dagger \rangle  -  \langle d_j^\dagger e_i(t) \rangle \Bigr) \\
 G^r_{cj}(t) &=& -i \theta(t) \Bigl( \langle c(t) d_j^\dagger \rangle  -  \langle d_j^\dagger c(t) \rangle \Bigr) \\
 G^r_{rj} (t)&=& -i \theta(t) \Bigl( \langle f(t) d_j^\dagger \rangle  -  \langle d_j^\dagger f(t) \rangle \Bigr)\\
 G^r_{pni,j}(t) & =& -i \theta(t) \Bigl( \langle d_i(t)(p^\dagger_i(t) +p_i(t)) d_j^\dagger \rangle  \\
& & -  \langle d_j^\dagger d_i(t)(p^\dagger_i(t) +p_i(t)) \rangle \Bigr) \ .
\end{array}
\label{GFencore}
\end{equation}
The first three of them can be calculated using equation of motion, and their FT lead to
\begin{equation}
\begin{array}{lcl}
G^r_{bij}(\omega) & = &  \frac{V_b}{\omega-E_i} G^r_{ij}(\omega)\\
G^r_{cj}(\omega) & = &\sum_l \frac{V_{cl}}{\omega-E_c} G^r_{lj}(\omega)\\
G^r_{rj}(\omega) & = &\sum_l \frac{V_{rl}}{\omega-E_r} G^r_{lj}(\omega) \ .
\end{array}
\end{equation}
The last one from Eq. (\ref{GFencore}) needs approximations to be evaluated, and will be expressed in terms of a self-energy reflecting exciton-phonon interaction (see later). 
The FT of Eq. (\ref{deriv}) can now be expressed as
\begin{equation}
\begin{array}{lcl}
\omega G^r_{ij}(\omega) &=&  \delta_{ij} + E_i G^r_{ij} (\omega) + \sum_l V_{il} G^r_{lj}(\omega)  \\
&+& g_i G^r_{pni,j}(\omega) + \sum_{\alpha=r,c,bi}  V_{\alpha i} G^r_{\alpha j}(\omega) \ .
\end{array}
\end{equation}
We finally rewrite it in a matrix notation, in the basis of sites, making the self energy apparent: 
\begin{equation}
G^r(\omega) = g^r(\omega) + g^r(\omega) \Bigl(V +\Sigma^r (\omega)\Bigr) G^r (\omega)\ ,
\label{Dyson}
\end{equation}
where $[g^{r}  (\omega)]_{ij}=\delta_{ij} / \bigl( \omega-E_i + i \eta \bigr) $ is the GF for the disconnected molecular site $i$.
Eq. (\ref{Dyson}) is known as Dyson equation. It makes an explicit reference to the $V_{ij}$ matrix but acquires a more familiar expression in the FMO mode basis:
\begin{equation}
G^r =\tilde{g}^r + \tilde{g}^r \Sigma^r G^r \ ,
\end{equation}
where $[\tilde{g}^{r}]_{kk'}  (\omega)=\delta_{kk'} / \bigl( \omega-\tilde{E}_k + i \eta \bigr) $, with $\tilde{E}_k$ the FMO energies, that is, the eigenvalues of the 
$H_{FMO}$ Hamiltonian (Eq. (\ref{HFMO})), and $\eta \rightarrow 0^+$.

The self-energy matrix can be split as
\begin{equation}
\Sigma^r = \Sigma_c^r + \Sigma_r^r + \Sigma_{b}^r+ \Sigma_{pn}^r \ ,
\label{Sigmar}
\end{equation}
with an exact first part
\begin{equation}
\begin{array}{lcl}
\Sigma_c^r (\omega) &= &g^r_c(\omega) W_c \\
\Sigma_r^r (\omega) &=&  g^r_r(\omega) W_r \\
 \bigl[\Sigma_{b}^r (\omega)\bigr]_{ij} &=& \delta_{ij} g^r_{bi}(\omega) V_b^2  \ ,
 \end{array}
 \label{Sigmar0}
\end{equation}
written in terms of $[W_{\alpha}]_{ij}=  V_{\alpha i}V_{\alpha j}$
for $\alpha=r,c$.
The second part of Eq. (\ref{Sigmar}), related to interaction with vibrations, needs approximations, that are detailed in the next Appendix and lead to
\begin{equation}
 \Sigma_{pn}^r   =   2  \lambda k_B T \Delta G^r \ .
\label{Sigmarpn}
\end{equation}
with $ \Delta$ a matrix defined as follows:
\begin{equation}
 \bigl[ \Delta A \bigr]_{ij} = \delta_{ij}A_{ij} \ .
 \label{MDelta}
\end{equation}
Due to this last term, which depends on $G^r$, Eq. (\ref{Dyson}) has to be self-consistently evaluated.

\subsubsection{$G^<_{ij}$ }

We now turn to the evaluation of $G^<_{ij}(\omega)$. Again NEGF, especially through Langreth rules, sets out the requirements to evaluate this function from Eq. (\ref{Dyson}). 
It gives
\begin{equation}
G^< =  G^r \bigl( (g^r)^{-1} g^< (g^a)^{-1} + \Sigma^< \bigr)G^a \ .
\label{Dysonless}
\end{equation}
In the steady state regime at hand, the GF may not depend of the population of the various levels that prevailed in the remote past when the different system parts were disconnected. We can therefore ignore the first term
to obtain the usual Keldysh equation, here in matrix form:
\begin{equation}
G^< =  G^r  \Sigma^< G^a  \ ,
\label{Keldysh}
\end{equation}
with 
$\Sigma^< = \Sigma_c^< + \Sigma_r^< + \Sigma_{b}^< + \Sigma_{pn}^<$. In details
\begin{equation}
\begin{array}{lcl}
\Sigma_c^< (\omega) &= &g^<_c(\omega) W_c= 2 i  n_c  \mathrm{Im} \bigl(g^r_c(\omega) \bigr) W_c  \\
\Sigma_r^< (\omega) &=& g^<_r(\omega) W_r=  2 i  n_r  \mathrm{Im} \bigl(g^r_r(\omega) \bigr) W_r\\
 \bigl[\Sigma_{b}^< (\omega)\bigr]_{ij} &=& \delta_{ij} V_b^2 g^<_{bi}(\omega) = 2 i  n_{bi} V_{b} \mathrm{Im} \bigl(g^r_{bi}(\omega) \bigr) V_{b} \ ,
 \label{sigless}
 \end{array}
\end{equation}
and 
\begin{equation}
\Sigma_{pn}^< = 2 \lambda k_B T \Delta G^< \ .
\label{Sigmalpn}
\end{equation}
The approximations leading to the previous expressions for $\Sigma_{pn}^r$ and $\Sigma_{pn}^<$ are detailed in the next paragraph.
 In the present treatment the exciton-phonon interaction is supposed elastic. 
Now, the Keldysh Eq. (\ref{Keldysh}) can be also self-consistently calculated. However it depends on $G^r$, thus the latter has to be computed first.

Before turning to the self-energy expression, it should be noted that knowing $G^<_{ij}(\omega)$ makes it possible to evaluate the local exciton density $n_i$ defined by
its integrated diagonal part  
\begin{equation}  
 n_i = i  \int \frac{d \omega}{2 \pi} G^<_{ii} (\omega)  \ .
\end{equation}

\section*{Appendix B: Self energy for exciton-vibration interaction}

In this paragraph we examine  the self-energy that characterizes  the exciton-vibration interaction. 
Numerous studies have focus on electron-phonon interaction in translation invariant systems. We draw inspiration from works in molecular or atomic junctions as those of Refs. 
[\onlinecite{Bihary05}] and [\onlinecite{Lu07}] where this invariance is lost. 
In the case at hand, for local exciton-phonon interaction whose Hamiltonian appears in Eq. (\ref{hepn}), the self-consistent Born approximation leads to the following retarded and lesser self-energies [\onlinecite{HaugJauho, Bihary05, Lu07}]
\begin{equation}
\begin{array}{lcl}
\bigl[ \Sigma^r_{pn}(\omega) \bigr]_{ij}& = & i g^2  \delta_{ij} \int \frac{d \nu}{2 \pi}  \Bigl( G_{ii}^<(\nu)D_i^r(\omega-\nu) \\
&+& G_{ii}^r(\nu)D_i^<(\omega- \nu) + G_{ii}^r(\nu) D_i^r(\omega-\nu)\Bigr) \\
& + &   \delta_{ij} g^2 n_i D^r(\omega'=0) \ , \\
\bigl[ \Sigma^<_{pn}(\omega)  \bigr]_{ij}& = & i g^2  \delta_{ij} \int \frac{d \nu}{2 \pi} G_{ii}^<(\nu) D_{i}^<(\omega-\nu)  \ , 
\end{array}
\label{Sigmapn}
\end{equation} 
with $n_i$ the exciton number at chromophore $i$, and
the phonon GF defined as follows [\onlinecite{HaugJauho}]:
\begin{equation}
D_i^r (\omega)= \frac{1}{\omega-\omega_{0i} + i \eta} - \frac{1}{\omega+\omega_{0i} + i \eta}
\end{equation}
\begin{equation}
D_i^< (\omega)= - 2 i \pi \Bigl[  N_i(\omega_{0i})  \delta (\omega-\omega_{0i}) +( N_i(\omega_{0i})  +1) \delta (\omega+\omega_{0i})\Bigr] 
\label{Dless}
\end{equation}
where $ N_i(\omega_{0i}) $ is the phonon number at chromophore $i$. 
Assuming a phonon bath at equilibrium, and in the high temperature regime for which $k_B T \gg \hbar \omega_{0i}$, the inequality $ N_i(\omega_{0i}) \gg 1$ allows to neglect the terms in 
$D^r_i$ compared to the one with $D^<_i$. 
We thus can write in a compact notation 
\begin{equation}
\bigl[ \Sigma^{r,<}_{pn}(\omega)  \bigr]_{ij} =  i g^2  \delta_{ij} \int \frac{d \nu}{2 \pi} G_{ii}^{r,<}(\nu) D_{i}^<(\omega-\nu)  \ .
\end{equation}
A further approximation, is usually made in Eq. (\ref{Dless}) in which the phonon frequencies are neglected compared with those of excitons, leading to
$ D_i^< (\omega) \simeq - 2 i \pi   \bigl( 2 N_i(\omega_{0i}) +1 \bigr)   \delta (\omega) $, which, with $N_i(\omega_{0i}) \simeq \frac{k_B T}{\hbar \omega_{0i}}$ leads to
\begin{equation}
\bigl[ \Sigma^{r,<}_{pn}(\omega)  \bigr]_{ij} =  2  g^2  \delta_{ij} \frac{k_B T}{\hbar \omega_{0i}} G_{ii}^{r,<}(\omega)  = 2  \delta_{ij} \lambda k_B T G_{ii}^{r,<}(\omega) \ .
\end{equation}
We have introduced the reorganization energy parameter $\lambda$ defined in the main text.
Using the $\Delta$ matrix previously defined in Eq. (\ref{MDelta}), we can gather the results of the present Appendix in the following form
\begin{equation}
 \Sigma^{r,<}_{pn}(\omega)  =  2   \lambda k_B T  \Delta G^{r,<}(\omega) \ .
\end{equation}

\section*{Appendix C: $\alpha$ Matrix }

In the present Appendix, we establish Eq. (\ref{linearite}) and evaluate the various coefficients of the $\alpha$ matrix. 

From Eqs. (\ref{Jinout}), (\ref{jialpha}),  (\ref{glessic}) and (\ref{GFlesshyb}), and the definition of self-energies, using the notation Tr for the trace of a matrix, we can rewrite the expressions already quoted in Eq. (\ref{Jcr}).  
\begin{equation}
\begin{array}{lcl}
J_c  & =& - \frac 2 \hbar \mathrm{Re} \int \frac{d \omega}{2 \pi} \mathrm{Tr} \bigl[ G^r \Sigma_c^< + G^< \Sigma^a_c \bigr] \\
J_r  & =& \frac 2 \hbar \mathrm{Re} \int \frac{d \omega}{2 \pi} \mathrm{Tr} \bigl[ G^r \Sigma_r^< + G^< \Sigma^a_r \bigr]  \ ,
\end{array}
\label{Jcrbis}
\end{equation}
where we have introduced the advanced reservoir self-energy, on the model of the retarded ones (Eq.(\ref{Sigmar0})):
\begin{equation}
\Sigma_\alpha^a (\omega)= W_\alpha g^a_\alpha(\omega) 
\label{sigmaa}
\end{equation} 
for $\alpha=r,c$.
Note that $\Sigma^a_\alpha$ are independent on the reservoir populations $n_c$ and $n_r$.
We can notice also that, according to Eq. (\ref{sigless}),  $\Sigma_c^< \propto n_c$, while $\Sigma_r^< \propto n_r$. We take this opportunity to define 
$\tilde{\Sigma}_c^<  = \Sigma_c^< /  n_c$, and $\tilde{\Sigma}_r^<  = \Sigma_r^< /  n_r$. 
We are going to demonstrate that $G^r$, in the high temperature hypothesis for the exciton-phonon self-energy, does not depend on neither  $n_c$ nor $n_r$, and that,
in a matrix notation,  
$G^< =  n_c G^r Y_c G^a+ n_r G^r Y_r G^a$, with $Y_c(\omega)$ and $Y_r(\omega)$, two expressions independent on $n_c$ and $n_r$, to be given soon.
Relying on Eq. (\ref{Jcr}), 
this will give rise to Eq. (\ref{linearite}), and enables to express the $\alpha$-matrix elements. 

First, it is easy to show that $G^r$ does not depend on the reservoir populations: it is evaluated from the Dyson equation, Eq. (\ref{Dyson}), in a self-consistent manner, that is, at stage $n+1$, the GF 
$G^{r(n+1)}$ is obtained by evaluating $g^r + g^r \Bigl(V +\Sigma^{r(n)} \Bigr) G^{r(n)}$.  
The self energy $\Sigma^r$, is evaluated in Eqs. (\ref{Sigmar}), (\ref{Sigmar0}) and (\ref{Sigmarpn}), with, 
in the high temperature limit (see previous Appendix), $\Sigma^{r(n)}_{pn} = 2 \lambda k_B T \Delta G^{r(n)}$.
This loop is repeated until convergence, and it is clear that, 
since $\Sigma_r^c$ and $\Sigma^r_r$ are independent of $n_c$ and $n_r$, 
the same applies to $G^r$. 
However this result relies on the high temperature hypothesis: indeed, except in this case, another term depending on $G^<$ remains in $\Sigma^r_{pn}$ (see Eq. (\ref{Sigmapn})).
And, as we shall prove soon, $G^<$ is an affine function of $n_r$ and $n_c$. 

We now turn to $G^<$. From Eqs. (\ref{Keldysh}), (\ref{sigless}) and (\ref{Sigmalpn}), it is easy to get 
\begin{equation}
\Sigma^<_{pn} = x \Delta \Bigl\{ G^r \bigl[ n_c \tilde{\Sigma}^<_c + n_r \tilde{\Sigma}^<_r + \Sigma^<_{pn} \bigr]  G^a  \Bigr\} \ ,
\end{equation}
where we have use the short-hand notation $x = 2 \lambda k_B T$. 
This equation can be expressed in a power expansion of $x$, with all terms proportional to $n_c$ or $n_r$. With the following notations, 
$  \Sigma^<_{pn} = X_c n_c + X_r n_r$, we get 
\begin{equation}
G^< = n_c G^r \bigl( \tilde{\Sigma}^<_c  +  X_c \bigr) G^a + n_r G^r  \Bigl( \tilde{\Sigma}^<_r + X_r \bigr) G^a \ ,
\end{equation}
from which the $\alpha$-matrix coefficients can now be obtained, as written in Eq. (\ref{alphacoeff}).    

By symmetry we expect $\alpha_{rc}= \alpha_{cr}$. This can be established easily for the first parts of $\alpha_{cr}$ and $\alpha_{rc}$, implying respectively 
$Re \bigl[ G^r \tilde{\Sigma}^<_r G^a \Sigma^a_c \Bigr]$ and $Re \bigl[ G^r \tilde{\Sigma}^<_c G^a \Sigma^a_r \Bigr]$. We first note that $G^a = \bigl( G^r \bigr)^*$.
Using Eq. (\ref{sigmaa}) and the one for $\Sigma^r_\alpha$, 
we get
\begin{equation}
Re \bigl[ G^r \tilde{\Sigma}^<_r G^a \Sigma^a_c \Bigr] = Re \bigl[ 2 i  \mathrm{Im} \bigl( g^r_r\bigr) g^a_c G^r W_r \bigl(G^r\bigr)^* W_c \Bigr]   \ .
\end{equation}
We need to examine $  Re \bigl[ 2 i  \mathrm{Im} \bigl( g^r_r\bigr) g^a_c\Bigr]  $. From the definitions of the reservoir advanced or retarded GF previously presented, it it easy to show that
$ Re \bigl[ 2 i  \mathrm{Im} \bigl( g^r_r\bigr) g^a_c\Bigr] = Re \bigl[ 2 i  \mathrm{Im} \bigl( g^r_c\bigr) g^a_r\Bigr] $, which establishes
\begin{equation}
Re \bigl[ G^r \tilde{\Sigma}^<_r G^a \Sigma^a_c \Bigr] =  Re \bigl[ G^r \tilde{\Sigma}^<_c G^a \Sigma^a_r \Bigr]  \ .
\end{equation}
The same argument can be repeated for the second parts of $\alpha_{cr}$ and $\alpha_{rc}$, for example, for each term in the expansion in power of $x$ for $X_r$ and $X_c$, leading to
\begin{equation} 
Re \bigl[ G^r X_r G^a \Sigma^a_c \Bigr] =  Re \bigl[ G^r X_c G^a \Sigma^a_r \Bigr]  \  ,
\end{equation}
and to the expected equality $\alpha_{cr} = \alpha_{rc}$.

\section*{Appendice D}
\begin{figure}
\includegraphics[width=0.5\textwidth]{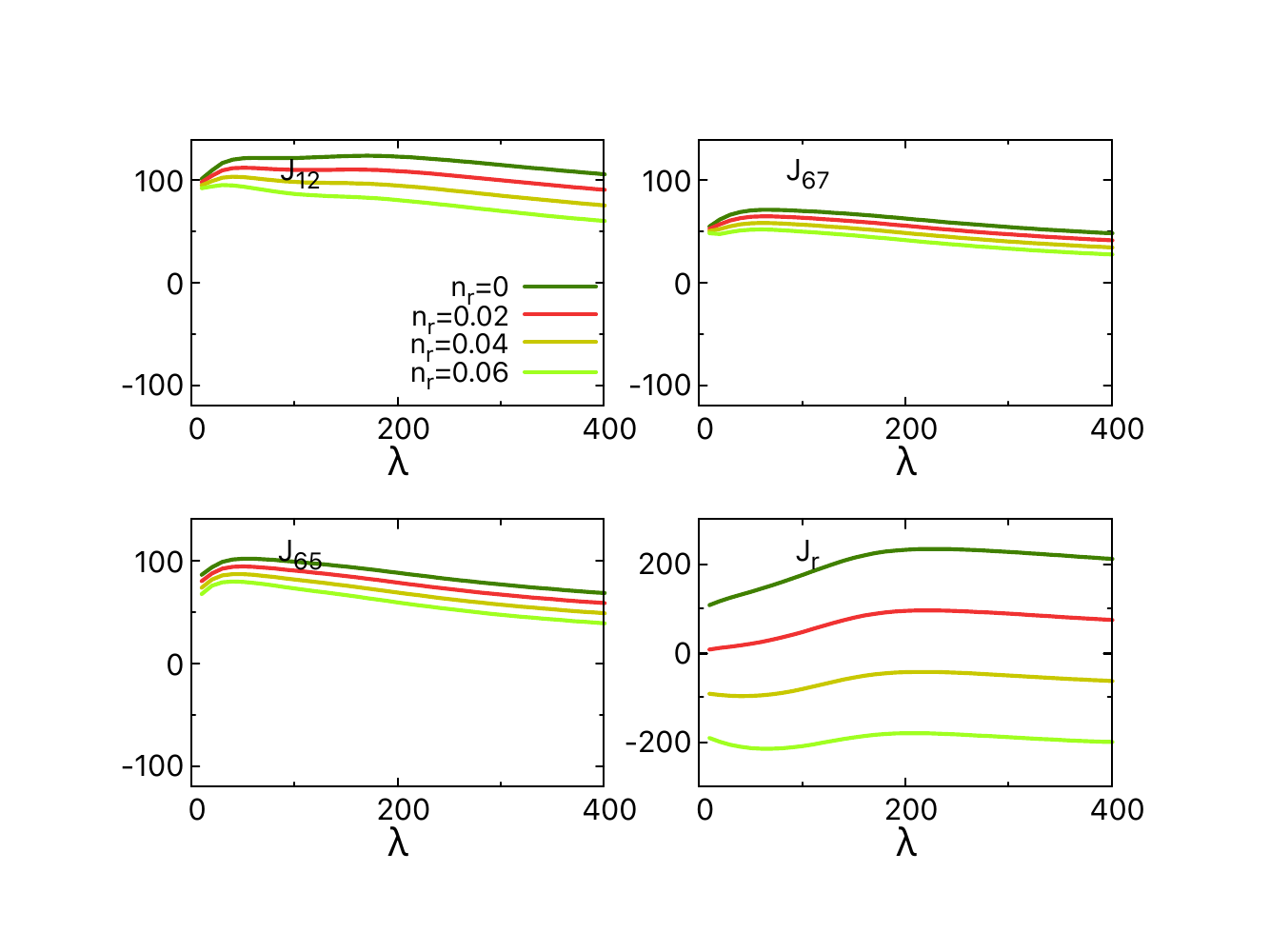}
\caption{  For four different $n_r$ values, different inter-BChl FMO currents as a function of $\lambda$, as well as output current, for $n_c=1$. 
}
\label{LocalCurrents}
\end{figure}

Fig.~\ref{LocalCurrents} completes Fig.~\ref{MainLocalCurrents} that was presented in the main text.
Due to a larger distance from the reaction center, these inter-BChl currents are less sensitive to the value of $n_r$.


\begin{thebibliography}{99}
\bibitem{remarque}  A misprint has probably slipped through in the paper [\onlinecite{Pelzer14}] concerning the $V_b$ value: indeed $V_b \tau \simeq h $ gives 
$V_b = 0.033$ cm$^{-1}$ for $\tau \simeq 1$ ns.
\bibitem{number} We are interested here in the non diagonal elements of the $\alpha$ matrix. This matrix being real and symmetric, this leads at most to 
$( 7 \times 7 -7)/2= 21 $ different off-diagonal elements.
\end{thebibliography}
\end{document}